\documentclass[aps,pra,epsfig,twocolumn,showpacs,superscriptaddress]{revtex4}

\def\be{\begin{equation}}        
\def\ee{\end{equation}}
\def\bea{\begin{eqnarray}}
\def\eea{\end{eqnarray}}
\def\bma{\begin{mathletters}}
\def\ema{\end{mathletters}}
\def\bi{\begin{itemize}}
\def\ei{\end{itemize}}

\def\C{\hbox{$\mit I$\kern-.7em$\mit C$}}

\newcommand{\ket}[1]{ | \, #1  \rangle}
\newcommand{\bra}[1]{ \langle #1 \,  |}
\newcommand{\proj}[1]{\ket{#1}\bra{#1}}

\begin{document}

\title{Locally Accessible Information and Distillation of Entanglement}

\author{Sibasish Ghosh}
\email{sibasish@imsc.res.in}
\affiliation{Institute of Mathematical Sciences, C. I. T. Campus, Taramani, Chennai 600113, India}

\author{Pramod  Joag}
\email{pramod@physics.unipune.ernet.in}
\affiliation{Department of Physics, University of Pune, Ganeshkhind, Pune 411 007, India}

\author{Guruprasad Kar}
\email{gkar@imsc.res.in}
\affiliation{Institute of Mathematical Sciences, C. I. T. Campus, Taramani, Chennai 600113, India}
\affiliation{Physics and Applied Mathematics Unit, Indian statistical Institute, 203, B. T. Road, Kolkata 700 108, India}

\author{Samir Kunkri}
\email{skunkri_r@isical.ac.in}
\affiliation{Physics and Applied Mathematics Unit, Indian statistical Institute, 203, B. T. Road, Kolkata 700 108, India}

\author{Anirban Roy}
\email{anirb@imsc.res.in}
\affiliation{Institute of Mathematical Sciences, C. I. T. Campus, Taramani, Chennai 600113, India}

\date{\today}

\begin{abstract}
A new type of complementary relation is found between locally accessible information and final average entanglement for given ensemble. It is also shown that in some well known distillation protocol, this complementary relation is optimally satisfied. We discuss the interesting trade-off between locally accessible information and distillable entanglement for some states.  

\end{abstract}


\maketitle

The problem of local distinguishability of orthogonal quantum states has raised much interest in the arena of quantum information. Interestingly where any two pure orthogonal states can be distinguished locally \cite{hardy00}, there exists more than two orthogonal states which can not be distinguished by local operations and classical communication (LOCC) \cite{ghosh01, horo03}. All these results lead to investigate the connection between the locally accessible information and amount of quantum entanglement of an ensemble. 

Recently Badzi{\c{a}}g \emph{et. al} \cite{bad03} has found a universal Holevo-like upper bound on the locally accessible information. This bound not only involves local entropy but also initial average entanglement. In particular they have shown that for an ensemble $\cal{E}$ $ = \{p_x, \rho_x\}$, the locally accessible information (information of $x,$ extractable by LOCC) is bounded by 
\begin{equation}
\label{uholevo}
I_{\rm acc}^{\rm LOCC} \le n - \overline{E}
\end{equation}
with $n = log_2d_1d_2$ for a $d_1 \otimes d_2$ system and $\overline{E}$ refers to any asymptotically consistent measure of the average entanglement of the ensemble. Now if one writes the inequality in the form $ I_{\rm acc}^{\rm LOCC} + \overline{E} \le n$, it shows some kind of complementarity relation between locally accessible information and the average entanglement. 
 Various interesting results follow from this relation. Specifically  Badzi{\c{a}}g \emph{et. al} \cite{bad03} has checked that given the dimensions of the systems what would be the ensemble that would saturate the bound. One can observe from this inequality that though there are extreme cases where $I_{\rm acc}^{\rm LOCC} = log_2d_1d_2$, there cannot be other extreme viz. $\overline{E} = log_2d_1d_2$, rather $\overline{E} \le {\rm min}\{log_2d_1, log_2d_2\}.$

We conjecture in this letter a modified inequality which involves not only the average entanglement of the initial ensemble ($\overline{E_i}$) but also the average entanglement of the final ensemble ($\overline{E_f}$). In other words, the amount of locally accessible information $I_{\rm acc}^{LOCC}$ is bounded above by $n - \overline{E_i} - \overline{E_f},$ which can be rewritten in the following form  
\begin{equation}
\label{modholevo}
I_{\rm acc}^{\rm LOCC} + \overline{E_f} \le n - \overline{E_i}
\end{equation}
Then for the given choice of ensemble (\emph{i. e.} for fixed $n$ and $\overline{E_i}$), there is kind of a complementarity between $I_{\rm acc}^{\rm LOCC}$ and the final average entanglement. 

We shall prove the inequality (\ref{modholevo}) for 1-way LOCC and will provide some simple examples to check the nice trade-off between amount of locally accessible information and final average entanglement in the above-mentioned complementarity relation and also notice how this have a role in the process of entanglement distillation. 

Finally we will discuss some famous distillation protocols like hashing, breeding and also error correcting protocol where our bound in inequality (\ref{modholevo}) saturates.

In the following we will provide a proof of inequality (\ref{modholevo}) for one way LOCC. 
When a source prepares a state $\rho_X$ where $X=0,....,n$ with
probabilities $p_{0},...,p_{n.}$ the Holevo bound tells us the maximal
accessible information (how well the source state can be inferred) one can derive
is bounded by the following limit. 
$$I_{\rm acc}\le S(\rho )-\sum_X p_{X}S(\rho _{X})$$
where $\rho =\sum_X p_{X}\rho _{X}$ and $S()$ is the von Neumann entropy.
But we will be considering a more interesting problem. The source is emitting
a bipartite state \( \rho ^{(AB)}_{X} \) where \( X=0,....,n \) with probabilities
\( p_{0},....,p_{n} \) and two particles $A$ and $B$ are given to two distant parties (Alice
and Bob, say) who are trying to guess \( X \) by LOCC. We will be trying to derive the upper limit of accessible
information by LOCC. As only LOCC is allowed so one among Alice and Bob has to start the
protocol. Let Alice starts it. Alice can look at it in the following manner.
She gets the states \( Tr_{B}[\rho ^{(AB)}_{X}] \) with probabilities \( p_{0},....,p_{n} \)
and she has to identify \( X. \) So she performs a measurement described by POVM
elements $\{ A_y \} = \{A_0, A_1, \dots, A_m \}$ on her system and by this process at the most what information she can extract about \( X \) is limited
by the Holevo bound, which is \[
I^{(A)} \leq S(\rho ^{(A)})-\sum_X p_{X}S(\rho ^{(A)}_{X})\]
 where \( \rho ^{(A)}=Tr_{B}[\rho ^{(AB)}]=Tr_{B}[\sum p_{X}\rho ^{(AB)}_{X}] \)
and \( \rho ^{(A)}_{X}=Tr_{B}[\rho ^{(AB)}_{X}]. \) The above inequality can
be rewritten in the following way. \[
I^{(A)} \leq log_{2}d_{1}-\sum_X p_{X}E(\rho ^{(AB)}_{X})=log_{2}d_{1}-\overline{E_{i}}\]
 where \( d_{1} \) is the dimension of the Hilbert space at Alice's side and
\( E \) is any asymptotic entanglement measure. Here we used the fact that \( S(\rho ^{(A)}_{X})\geq E(\rho ^{(AB)}_{X}) \)
for any asymptotic entanglement measure \( E \). We will define \( \overline{E_i} = \sum_X p_X E(\rho_X^{AB}) \)  as initial average entanglement. After Alice's extraction of information she
communicates her result, say $K$ to Bob. The joint two-particle density matrix has transformed into
\[ \frac{[A_K \otimes I \sum_X p_{X}\rho _{X}^{(AB)} A_K^{\dag} \otimes I]}{Tr (A_K \otimes I \sum_X p_{X}\rho _{X}^{(AB)} A_K^{\dag} \otimes I)}  = \sigma_{K}^{(AB)} \] with probability $p_K = Tr (A_K \otimes I \sum_X p_{X}\rho _{X}^{(AB)} A_K^{\dag} \otimes I).$ 
 And then Bob's state is transformed into $\sigma_K ^{(B)}=Tr_{A} \sigma_{K}^{(AB)}$ (which includes information of $X$ accessible by Bob) with probability $p_K.$ 

Now we will be using some more notations. We define  \[ \sigma_{KX}^{(AB)} = \frac{[A_K \otimes I \rho _{X}^{(AB)} A_K^{\dag} \otimes I]}{Tr (A_K \otimes I \rho _{X}^{(AB)} A_K^{\dag} \otimes I)} \] and $p_{KX} = Tr[A_k \otimes I \rho _{X}^{(AB)} A_K^{\dag} \otimes I].$  

It's now Bob's turn to perform the measurement depending on Alice's outcome (here $K$. say) to extract information of $X.$ He performed a measurement with POVM elements $\{ B_z \} = \{B_0, B_1, \dots B_l \}$ on his system to extract information about $X$ on the ensemble \[ \sigma_K ^{(B)} = \sum_X \frac{p_X p_{KX}}{p_K} Tr_A \sigma_{KX}^{(AB)}, \] which originated from Alice's $K$th measurement outcome. The accessible information of Bob $I^{(B)}_{K}$ must be bounded above by the Holevo quantity $S(\sigma_K ^{(B)}) - \sum_X p^{\prime}_{KX} S(Tr_{A} \sigma_{KX}^{(AB)})$ where $p^{\prime}_{KX} = (p_X p_{KX})/p_K.$ Thus we have
\[ I^{(B)}_{K} \leq S(\sigma_K ^{(B)}) - \sum_X p^{\prime}_{KX} S(Tr_{A} \sigma_{KX}^{(AB)})  \]
\[ = S(\sigma_K ^{(B)}) -  \sum_X p^{\prime}_{KX} S \left( Tr_{A} \left( \sum_i \lambda_i^{KX} \proj{\psi_i^{KX}} \right) \right) \]
\[ \le  S(\sigma_K ^{(B)}) - \sum_X p^{\prime}_{KX} \sum_i \lambda_i^{KX} S \left( Tr_A \proj{\psi_i^{KX}} \right) \] (by using concavity of von Neumann entropy), 
\[ = S(\sigma_K ^{(B)}) - \sum_X p^{\prime}_{KX} \sum_i \lambda_i^{KX} E \left( \proj{\psi_i^{KX}} \right) \] 
\[ \le S(\sigma_K ^{(B)}) - E_F ( \sigma_K ^{(B)} ) \] (by definition of entanglement of formation $E_F$)
\[ \le S(\sigma_K ^{(B)}) - E ( \sigma_K ^{(B)}) \]  (where E is any asymptotic measure of entanglement and is smaller than the entanglement of formation)
\[ \le log_{2}d_{2} - E ( \sigma_K^{(B)}) \]

 where $\sum_i \lambda_i^{KX} \proj{\psi_i^{KX}}$ is any decomposition of $ \sigma_{KX}^{(AB)},$ and $log_2 d_2$ is the dimension of Hilbert space at Bob's side.   

The total bound on Bob's extractable information, is \( I^{(B)} \) where 
 \( I^{(B)} \le \sum_K p_K I^{(B)}_K, \) which can be rewritten as 

$$I^{(B)} \leq \sum_K p_K log_{2}d_{2} - \sum_K p_K E ( \sigma_K^{(B)} ).$$

where $ \sum_K p_K E ( \sigma_K^{(B)} ) $ is the average entanglement before Bob's measurement. Let $\overline{E_f}$ is the final average entanglement after Bob's measurement and as
average entanglement can only decrease by LOCC so $$I^{(B)} \leq \sum_K p_K log_{2}d_{2} - \sum_K p_K E ( \sigma_K^{(B)} ) \le log_2 d_2 - \overline{E_f}.$$

So in this 1-way protocol
the total locally accessible information satisfies the following relation 

\begin{equation}
\label{lholevo}
I_{\rm acc}^{\rm LOCC} \le I^{(A)} + I^{(B)} \leq log_{2}d_{1}+log_{2}d_{2}-\overline{E_{i}}-\overline{E_{f}}
\end{equation}
 
Hence the complementarity relation has been established between locally accessible information and final average entaglement for given ensemble. Now in some special cases, if all or some of the component states of the final ensemble generated by the LOCC are maximally entangled states, the process of extraction of ensemble information (locally) has also distilled some entanglement. Obviously the amount of entanglement ($E_{\rm Distilled}$) that may be distilled in this process will satisfy $ E_{\rm Distilled} \le \overline{E_f}.$   
So for every distillation process, we can also present a complementarity relation as follows
$$ I_{\rm acc}^{\rm LOCC} + E_{\rm Distilled} \le log_{2}d_{1}d_2-\overline{E_{i}}$$
If for some cases, $E_{\rm Distilled} = E_d$ (distillable entanglement) then this process of extraction of locally accessible information is itself the best distillation process.
 
First we will provide some simple examples (of course avoiding those discussed elsewhere \cite{ghosh01}) to find the implication of our inequality (inequality (\ref{modholevo})).\\ 

(\textbf{Ex. 1}) Consider the following example where the source is producing any one of the state $\rho_X$, which is  three copies of Bell sates, $X =1, 2, 3, 4,$ with probability $p_X=1/4,$ \emph{i.e.} Alice and Bob have the following ensemble $\cal{E}$ = $\{p_X=1/4, \rho_X=(\proj{B_X})^{\otimes 3}\}$. Here $\ket{B_X}$ are known Bell states $\ket{B_1} = \frac{1}{\sqrt{2}} \left(\ket{00}+\ket{11}\right),~ \ket{B_2} = \frac{1}{\sqrt{2}} \left(\ket{00}-\ket{11}\right), ~ \ket{B_3} = \frac{1}{\sqrt{2}} \left(\ket{01}+\ket{10}\right),~ \ket{B_4} = \frac{1}{\sqrt{2}} \left(\ket{01}-\ket{10}\right).$ Now the maximum amount of information about $X$ one can extract locally (or globally also) is $2$ cbit (\emph{i.e.} $I_{\rm acc}^{\rm LOCC} = 2$). Hence final average entanglement $\overline{E_f}$ is bounded above by $ log_2d_1d_2 - \overline{E_i} - I_{\rm acc}^{\rm LOCC} = 6 - 3 -2 = 1.$ 
By using two copies of the Bell states one can know the Bell state and therefore with the remaining copy, finally one can distill $1$ ebit. But there is a process given by Chen \emph{et. al.} \cite{chen03} by which one can extract $2$ ebits, then our inequality (inequality (\ref{modholevo})) shows that extractable information $I_{\rm acc}^{\rm LOCC}$  is bounded by $1.$ Using inequality (11) of Chen \emph{et. al.} \cite{chen03} one can easily check that $I_{\rm acc}^{\rm LOCC} = 1$ which also saturates our bound \cite{foot1}.

One can generalize this process for $n$ copies of Bell states, where $n$ is odd \emph{i.e.} the source is producing a state which is $n$ copies of one of the four Bell states with probability 1/4. The distillable entanglement is $(n-1)$ ebit \cite{chen03}. When one distills this amount of entanglement, $I_{\rm acc}^{\rm LOCC}$ can be at most $1$ cbit (from (\ref{modholevo})). But if one tries to extract maximum amount of classical information about the ensemble, \emph{i.e.} 2 cbit, the amount of entanglement one can distill is at most $(n-2)$ ebit. This can be achieved by using two copies of Bell states for reliable discrimination and rest $(n-2)$ copies produce $(n-2)$ ebit. \\

(\textbf{Ex. 2})  Another interesting example is $\{p_X=1/4,~ \rho_X=(\proj{B_X})^{\otimes 4} \},~X=1, \dots, 4.$ Here $log_2d_1d_2 = 8$, $\overline{E_i} = 4$, maximum allowed value of $I_{\rm acc}^{\rm LOCC}$ is $2$; and hence $\overline{E_f}$ is bounded above by $2$ which is equal to the entanglement one can distill by locally discriminating four Bell states using two copies. For this ensemble the distillable entanglement $E_D$ is also $2$ \cite{chen03}. 

For all even cases, as distillable entanglement is $(n-2)$ \cite{chen03}, here the extraction of  full $2$ bits of classical information is the best distillation process unlike the odd case.\\ 

(\textbf{Ex. 3}) We now take examples in $3 \otimes 3$ system. Take two copies of all $9$ maximally entangled states each of which are in the canonical form given by Eq. (\ref{mes3}) below with equal probability. 
\begin{equation}
\label{mes3}
 \ket{\Phi_{nm}^{(3)}} = \frac{1}{\sqrt{3}}\sum_{j=0}^{2}exp[\frac{2\pi ijn }{3}] \ket{j} \otimes \ket{(j+m) {\rm mod~ 3}}
\end{equation}
where $n, m = 0, 1, 2$ Here $n = log_2 d_1 d_2 = log_2 81 = 4log_2 3$, $ \overline{E_i}$ is $2 log_2 3$. So if $I_{\rm acc}^{\rm LOCC}$ is $2 log_2 3$ (which is maximum), $\overline{E_f}$ is $0.$ In another possibility, $\overline{E_f}$ can become $log_2 3.$ Then from our inequality (inequality (\ref{modholevo})), $I_{\rm acc}^{\rm LOCC}$ is $\le log_2 3.$ In a case where $I_{\rm acc}^{\rm LOCC} = log_2 3$ we show that the amount of entanglement that can be distilled is $log_2 3.$ We know from the works of Yang \emph{et. al.} \cite{yang04} the amount of distillable entanglement of $\rho_{3}^{(2)} = (1/9) {\sum}_{n, m = 0}^{2} (\proj{\Phi_{nm}})^{\otimes 2}$ is $log_2 3.$ Applying bilateral C-NOT operation  the ensemble $\rho_{3}^{(2)}$ transforms into $\frac{1}{3}[\proj{\Phi_{00}} \otimes \sum_n \proj{\Phi_{0n}} + \proj{\Phi_{10}} \otimes \sum_n \proj{\Phi_{2n}} + \proj{\Phi_{20}} \otimes \sum_n \proj{\Phi_{1n}}].$ Now one discriminates between subspaces spanned by $\{\ket{00}, \ket{11}, \ket{22}\}$, $\{\ket{01}, \ket{12}, \ket{20}\}$ and $\{\ket{02}, \ket{10}, \ket{21}\}$ and extracts  $I_{\rm acc}^{\rm LOCC} = log_2 3$ and at the same time distills $log_2 3$ ebit entanglement.

All the above examples show that whenever the state has some distillable entanglement, some amount of entanglement may be distilled in the process of extracting information about the ensemble. In some cases like \textbf{Ex. 1} and \textbf{Ex. 3} extracting full information about the ensemble reduces the amount to be distilled, but if one extracts some less information, the amount to be distilled reaches the distillable entanglement.

We now turn into $d \otimes d$ system. In $d \otimes d$, there are $d^2$ no of pairwise orthogonal maximally entangled states which can be written as $ \ket{\Phi_{nm}^{(d)}} = \frac{1}{\sqrt{d}}\sum_{j=0}^{d}exp[\frac{2\pi ijn }{d}] \ket{j} \otimes \ket{(j+m) {\rm mod~ d}}, n, m = 0, \dots ,d-1.$ These states can be discriminated either by providing two copies of each states \cite{ghosh02} or by sharing an additional amount of $log_2 d$ ebit of entanglement \cite{yang03}. These also follows from our bound. We also show that after having classical information no entanglement will remain finally.  As of previous example here $I_{\rm acc}^{\rm LOCC} = 2 log_2 d, n = 4 log_2 d, \overline{E_i} = 2 log_2 d$. So $\overline{E_f} \le 0.$ So finally no entanglement is there. Similarly for the case when one copy is supplied and $log_2 d$ amount of entanglement is also supplied, \emph{i.e.} a known maximally entangled state in $d \otimes d$ is supplied, the final average entanglement becomes zero after discrimination.     

Next we shall study the inequality (\ref{modholevo}) in the context of some famous distillation process like hashing, breeding and error correction protocol.  In distillation protocol like hashing, breeding the main idea is same as the classical problem of identifying a word for a given probability distribution of the alphabets which constitute the word. 

In the breeding protocol \cite{bennett96, wolf03}, sufficiently large no. of copies of Bell diagonal state $\rho_{B} = \sum_{i=1}^{4} p_i \proj{B_i}$ with corresponding Shannon entropy (\emph{i.e.} $H(p_i) = -\sum_i p_i log_2 p_i$)  less than $1$,  are considered. We are also supplied with $n H(p_i)$ no. of copies of a predistilled maximally entangled state. The $n$ copies of the Bell states ($\rho_B$) form the following string $\proj{B_{i_1}} \otimes \proj{B_{i_2}} \otimes \proj{B_{i_3}} \dots \otimes \proj{B_{i_n}}$ with probability $p_{i_1} p_{i_2} p_{i_3} \dots p_{i_n}$, and our job is to identify this string by using the predistilled states. So finally one gets $n$ no. of maximally entangled states. For this problem  $I_{\rm acc}^{\rm LOCC} = n H(p_i)$ (as the total no of different strings like $\proj{B_{i_1}} \otimes \proj{B_{i_2}} \otimes \proj{B_{i_3}} \dots \otimes \proj{B_{i_n}}$, that can be identified by the protocol, is $2^{nH(p_i)}$)   $\overline{E_i} = n(1+H(p_i)), ~~~ log_2 d_1 d_2 = 2n(1+ H(p_i)), ~~~ \overline{E_f} = n.$ Here one can see  the saturation of the bound given in inequality(\ref{modholevo}), in the asymptotic limit. 

In the hashing protocol \cite{bennett96, wolf03}, the string is identified, or equivalently the classical information is extracted at the expenses of the entanglement from the string. Starting with n copies of $\rho_B$ one gets $n (1 - H(p_i))$ copies of a known maximally entangled state in the asymptotic limit, by locally distinguishing $2^{n H(p_i)}$ no. of likely strings $\proj{B_{i_1}} \otimes \proj{B_{i_2}} \otimes \proj{B_{i_3}} \dots \otimes \proj{B_{i_n}}$ of the four Bell states, in which again our bound (\ref{modholevo}) saturates. 

In this context, our inequality establishes the fact that when distilling 
   from a mixture of Bell states $\sum_{i = 1}^{4} p_i \left|B_i\right\rangle 
   \left\langle B_i \right|$, if the process is to identify the strings of 
   Bell states in the ensemble (e.g., in breeding, hashing) by 1-way, or even 2-way LOCC, the highest amount of entanglement that can be distilled from 
   each copy of the Bell mixture is $1 - H(p_i)$. 

We are now going to discuss the relation between our bound and entanglement distillation by error correction. Let Alice and Bob shares $n$ non-maximally entangled states (they need not be the same), which arise due to the possible corruption during transmission of maximally entangled state from Alice to Bob by some noisy channel. Let the errors that occurred during the transmission belong to a subset, say $S$, of the Pauli group $G_n$ on $n$ qubits \cite{nielsenbook}, and there exists a stabilizer code to correct the errors \cite{shor00, lo99}. After the transmission one can write the $2n$ qubit state along with the environment as 
$$ \ket{\Psi_{A B E}} = \sum_i (I_A \otimes (U_i)_{B}) \ket{B_1}^{\otimes n} \ket{e_i}$$
where $\ket{e_i}$ are environment states (possibly non-orthogonal and unnormalized). Here $\{U_i\}$ is the set of unitary operators acting on the $2^n$ dim. Hilbert space of Bob's system, where each $U_i$ belongs to $S$, that can be corrected by the stabilizer code (characterized by ($n,m$)), considered in the problem. So the no. of linearly independent $U_i$'s are $2^{n-m}$. Now in this protocol Alice and Bob performs identical $(n-m)$-generator measurement on $n$ qubits in their possession, and comparing their measurement results they identify the error syndrome $i$ and then correct it. But in this process of measurement joint state of Alice and Bob collapsed to a maximally entangled entangled state of $2^m \otimes 2^m$. So finally Alice and Bob come up with an $m$ ebit maximally entangled state. 

As no knowledge of environment is used, this problem is equivalent to the problem of distilling maximally entangled state from the mixture 
\begin{equation}
\label{error}
\rho = \frac{1}{2^{n-m}} \sum_i (I_A \otimes ({U_i})_{B}) \left(\proj{B_1}\right)^{\otimes n} (I \otimes ({U_i})^{\dag}_{B})
\end{equation}
Thus in this process, the amount of information ($I_{\rm acc}^{\rm LOCC}$) that has been extracted is $(n-m)$ (by this process of error correction, we are detecting and then correctiong $2^{n-m}$ no. of equally probable errors appeared in (\ref{error}) by LOCC), $\overline{E_i} = n, ~~ \overline{E_f} = m, ~~ log_2 d_1 d_2 = 2n.$ So the bound (\ref{modholevo}) is saturated for this distillation protocol.

In this letter, we provided a relation (inequality (\ref{modholevo})) among 
   locally accessible information, initial average entanglement and final 
   average entanglement for any given asymptotic measure of entanglement. We 
   have given a proof of this relation for any 1-way LOCC and provided some 
   examples, each of which saturates the above-mentioned relation, revealing 
   complementarity between locally accessible information and the amount of 
   entanglement that has been distilled in this process. We have also shown 
   that in each of the three well-known distillation protocols -- breeding, 
   hashing, and distillation by error correction -- the above-mentioned 
   relation is saturated. Though all our examples (given here) involve 1-way 
   protocols, one can easily check that the inequality (\ref{modholevo}) is 
   strictly satisfied in the case of recurrence protocol \cite{bennett96a}.

   Distilling maximally entangled states from a general mixed state (created 
   due to some disturbance in the channel) by LOCC is a fundamental problem 
   in quantum information processing. Till now, the standard distillation 
   protocols deal with mixtures of Bell states, and in each of these 
   protocols, either full or partial (e.g., recurrence protocol) extraction 
   of information about the ensemble is performed. In particular, when for a 
   state, hashing and breeding protocols yield either no or very little 
   entanglement, initially recurrence protocol is used, in which partial 
   information about the ensemble is extracted to increase the fidelity. This 
   shows that extraction of full information about the ensemble may reduce 
   the amount of entanglement to be distilled. We have also encountered here 
   some examples where accessing full information about the ensemble, 
   distilled less amount than the corresponding distillable entanglement. All 
   these suggest that in order to find a better distillation protocol, one has to 
   take care about the interplay between the amount of accessible information 
   (to be accessed locally) and final average entanglement (which may equal to 
   the amount of entanglement distilled in the process), and optimize it in 
   some clever way. 

S. K. acknowledges the support by the Council of Scientific and Industrial Research, Government of India, New Delhi.


\begin{thebibliography}{14}
\expandafter\ifx\csname natexlab\endcsname\relax\def\natexlab#1{#1}\fi
\expandafter\ifx\csname bibnamefont\endcsname\relax
  \def\bibnamefont#1{#1}\fi
\expandafter\ifx\csname bibfnamefont\endcsname\relax
  \def\bibfnamefont#1{#1}\fi
\expandafter\ifx\csname citenamefont\endcsname\relax
  \def\citenamefont#1{#1}\fi
\expandafter\ifx\csname url\endcsname\relax
  \def\url#1{\texttt{#1}}\fi
\expandafter\ifx\csname urlprefix\endcsname\relax\def\urlprefix{URL }\fi
\providecommand{\bibinfo}[2]{#2}
\providecommand{\eprint}[2][]{\url{#2}}

\bibitem[{\citenamefont{Walgate et~al.}(2000)\citenamefont{Walgate, Short,
  Hardy, and Vedral}}]{hardy00}
\bibinfo{author}{\bibfnamefont{J.}~\bibnamefont{Walgate}},
  \bibinfo{author}{\bibfnamefont{A.~J.} \bibnamefont{Short}},
  \bibinfo{author}{\bibfnamefont{L.}~\bibnamefont{Hardy}}, \bibnamefont{and}
  \bibinfo{author}{\bibfnamefont{V.}~\bibnamefont{Vedral}},
  \bibinfo{journal}{Phys. Rev. Lett.} \textbf{\bibinfo{volume}{85}},
  \bibinfo{pages}{4972} (\bibinfo{year}{2000}).

\bibitem[{\citenamefont{Ghosh et~al.}(2001)\citenamefont{Ghosh, Kar, Roy, (De),
  and Sen}}]{ghosh01}
\bibinfo{author}{\bibfnamefont{S.}~\bibnamefont{Ghosh}},
  \bibinfo{author}{\bibfnamefont{G.}~\bibnamefont{Kar}},
  \bibinfo{author}{\bibfnamefont{A.}~\bibnamefont{Roy}},
  \bibinfo{author}{\bibfnamefont{A.~S.} \bibnamefont{(De)}}, \bibnamefont{and}
  \bibinfo{author}{\bibfnamefont{U.}~\bibnamefont{Sen}},
  \bibinfo{journal}{Phys. Rev. Lett.} \textbf{\bibinfo{volume}{87}},
  \bibinfo{pages}{277902} (\bibinfo{year}{2001}).

\bibitem[{\citenamefont{Horodecki et~al.}(2003)\citenamefont{Horodecki, (De),
  Sen, and Horodecki}}]{horo03}
\bibinfo{author}{\bibfnamefont{M.}~\bibnamefont{Horodecki}},
  \bibinfo{author}{\bibfnamefont{A.~S.} \bibnamefont{(De)}},
  \bibinfo{author}{\bibfnamefont{U.}~\bibnamefont{Sen}}, \bibnamefont{and}
  \bibinfo{author}{\bibfnamefont{K.}~\bibnamefont{Horodecki}},
  \bibinfo{journal}{Phys. Rev.Lett.} \textbf{\bibinfo{volume}{90}},
  \bibinfo{pages}{047902} (\bibinfo{year}{2003}).

\bibitem[{\citenamefont{Badzi{\c{a}}g et~al.}(2003)\citenamefont{Badzi{\c{a}}g,
  Horodecki, Sen(De), and Sen}}]{bad03}
\bibinfo{author}{\bibfnamefont{P.}~\bibnamefont{Badzi{\c{a}}g}},
  \bibinfo{author}{\bibfnamefont{M.}~\bibnamefont{Horodecki}},
  \bibinfo{author}{\bibfnamefont{A.}~\bibnamefont{Sen(De)}}, \bibnamefont{and}
  \bibinfo{author}{\bibfnamefont{U.}~\bibnamefont{Sen}},
  \bibinfo{journal}{Phys. Rev. Lett.} \textbf{\bibinfo{volume}{91}}
  (\bibinfo{year}{2003}).

\bibitem[{\citenamefont{Chen et~al.}(2003)\citenamefont{Chen, Jin, and
  Yang}}]{chen03}
\bibinfo{author}{\bibfnamefont{Y.-X.} \bibnamefont{Chen}},
  \bibinfo{author}{\bibfnamefont{J.-S.} \bibnamefont{Jin}}, \bibnamefont{and}
  \bibinfo{author}{\bibfnamefont{D.}~\bibnamefont{Yang}},
  \bibinfo{journal}{Phys. Rev. A} \textbf{\bibinfo{volume}{67}},
  \bibinfo{pages}{014302} (\bibinfo{year}{2003}).

\bibitem{foot1} Here both the parties applies C-NOT operations on their respective three qubit systems, transforming  $\rho^{(3)}$ to $\frac{1}{2}\left(\proj{B_1^{\otimes 2}}(\proj{B_1}+\proj{B_2})\right)+\frac{1}{2}\left(\proj{B_3^{\otimes 2}}(\proj{B_3}+\proj{B_4})\right)$ (see \cite{chen03}). One now distinguishes between whether the target state belongs to $\{\ket{00}, \ket{11}\}$ subspace or $\{\ket{01}, \ket{10}\}$ subspace. which is extracting 1 cbit and after this discrimination one distills 2 ebit.

\bibitem[{\citenamefont{Yang and Chen}(2004)}]{yang04}
\bibinfo{author}{\bibfnamefont{D.}~\bibnamefont{Yang}} \bibnamefont{and}
  \bibinfo{author}{\bibfnamefont{Y.-X.} \bibnamefont{Chen}},
  \bibinfo{journal}{Phys. Rev. A} \textbf{\bibinfo{volume}{69}},
  \bibinfo{pages}{024302} (\bibinfo{year}{2004}).

\bibitem[{\citenamefont{Ghosh et~al.}(2002)\citenamefont{Ghosh, Kar, Roy, and
  Sarkar}}]{ghosh02}
\bibinfo{author}{\bibfnamefont{S.}~\bibnamefont{Ghosh}},
  \bibinfo{author}{\bibfnamefont{G.}~\bibnamefont{Kar}},
  \bibinfo{author}{\bibfnamefont{A.}~\bibnamefont{Roy}}, \bibnamefont{and}
  \bibinfo{author}{\bibfnamefont{D.}~\bibnamefont{Sarkar}},
  \bibinfo{journal}{quant-ph/0205105}  (\bibinfo{year}{2002}).

\bibitem[{\citenamefont{Yang and Chen}(2003)}]{yang03}
\bibinfo{author}{\bibfnamefont{D.}~\bibnamefont{Yang}} \bibnamefont{and}
  \bibinfo{author}{\bibfnamefont{Y.-X.} \bibnamefont{Chen}},
  \bibinfo{journal}{quant-ph/0311100}  (\bibinfo{year}{2003}).

\bibitem[{\citenamefont{Bennett
  et~al.}(1996{\natexlab{a}})\citenamefont{Bennett, DiVincenzo, Smolin, , and
  Wootters}}]{bennett96}
\bibinfo{author}{\bibfnamefont{C.~H.} \bibnamefont{Bennett}},
  \bibinfo{author}{\bibfnamefont{D.~P.} \bibnamefont{DiVincenzo}},
  \bibinfo{author}{\bibfnamefont{J.~A.} \bibnamefont{Smolin}}, ,
  \bibnamefont{and} \bibinfo{author}{\bibfnamefont{W.~K.}
  \bibnamefont{Wootters}}, \bibinfo{journal}{Phys. Rev. A}
  \textbf{\bibinfo{volume}{54}}, \bibinfo{pages}{3824}
  (\bibinfo{year}{1996}{\natexlab{a}}).

\bibitem[{\citenamefont{Vollbrecht and Wolf}(2003)}]{wolf03}
\bibinfo{author}{\bibfnamefont{K.~G.~H.} \bibnamefont{Vollbrecht}}
  \bibnamefont{and} \bibinfo{author}{\bibfnamefont{M.~A.} \bibnamefont{Wolf}},
  \bibinfo{journal}{Phys. Rev. A} \textbf{\bibinfo{volume}{67}},
  \bibinfo{pages}{012303} (\bibinfo{year}{2003}).

\bibitem{nielsenbook} M. A. Nielsen, and I. L. Chuang, \emph{Quantum Computation and Quantum Information} (Cambidge University Press, Cambridge, UK, 2000) 

\bibitem[{\citenamefont{Shor and Preskill}(2000)}]{shor00}
\bibinfo{author}{\bibfnamefont{P.~W.} \bibnamefont{Shor}} \bibnamefont{and}
  \bibinfo{author}{\bibfnamefont{J.}~\bibnamefont{Preskill}},
  \bibinfo{journal}{Phys. Rev. Lett.} \textbf{\bibinfo{volume}{85}},
  \bibinfo{pages}{441} (\bibinfo{year}{2000}).

\bibitem[{\citenamefont{Lo and Chau}(1999)}]{lo99}
\bibinfo{author}{\bibfnamefont{H.-K.} \bibnamefont{Lo}} \bibnamefont{and}
  \bibinfo{author}{\bibfnamefont{H.~F.} \bibnamefont{Chau}},
  \bibinfo{journal}{Science} \textbf{\bibinfo{volume}{283}},
  \bibinfo{pages}{2050} (\bibinfo{year}{1999}).

\bibitem[{\citenamefont{Bennett
  et~al.}(1996{\natexlab{b}})\citenamefont{Bennett, Brassard, Poescu,
  Schumacher, Smolin, , and Wootters}}]{bennett96a}
\bibinfo{author}{\bibfnamefont{C.~H.} \bibnamefont{Bennett}},
  \bibinfo{author}{\bibfnamefont{G.}~\bibnamefont{Brassard}},
  \bibinfo{author}{\bibfnamefont{S.}~\bibnamefont{Poescu}},
  \bibinfo{author}{\bibfnamefont{B.}~\bibnamefont{Schumacher}},
  \bibinfo{author}{\bibfnamefont{J.~A.} \bibnamefont{Smolin}}, ,
  \bibnamefont{and} \bibinfo{author}{\bibfnamefont{W.~K.}
  \bibnamefont{Wootters}}, \bibinfo{journal}{Phys. Rev. Lett.}
  \textbf{\bibinfo{volume}{76}}, \bibinfo{pages}{722}
  (\bibinfo{year}{1996}{\natexlab{b}}).

\end{thebibliography}

\end{document}